\documentclass[doublespacing]{elsart}

\usepackage{graphicx}
\usepackage{amssymb}

\begin{document}
\begin{frontmatter}


\journal{SCES'2001}


\title{The Kondo lattice model from strong-coupling viewpoint}

%
%
%
%
%
%

\author{Valeri N. Kotov\corauthref{1}} 
\author{and Peter J. Hirschfeld}

%
 
\address{Department of Physics, University of Florida, Gainesville, 
FL 32611-8440, USA}

%
%
%
%


%
%
%
%

\corauth[1]{Corresponding Author: University of Florida, Department of Physics,
Gainesville, FL 32611-8440, USA. Fax: (352) 392-0524.
\newline
 E-mail: valeri@phys.ufl.edu,
 Valeri.Kotov@ipt.unil.ch (current)}


\begin{abstract}
We study the  magnetic  excitation spectrum  of
 the two-dimensional (2D) square-lattice S=1/2 Kondo lattice model at
finite (hole) doping, by  representing the Hamiltonian
 in terms of a set of local excitations.  
 The location of the paramagnetic-antiferromagnetic phase boundary at T=0 is
 determined.
\end{abstract}

%
%

\begin{keyword}
Quantum transitions \sep Kondo effect \sep Bond operators.  

\end{keyword}


\end{frontmatter}

%
%
%
%
%

 The Kondo lattice model in its  simplest form
 describes itinerant ($c$) electrons hopping on a lattice, and
 interacting with localized ($d$) orbitals via  Kondo
 exchange $J$ (we assume an antiferromagnetic sign $J\geq 0$):
\begin{equation}
H = -t \sum_{i,j,\sigma}  \left(
c_{\sigma,i}^{\dagger}c_{\sigma,j} +
\mbox{h.c.} \right) +J\sum_{i}
{\bf S}_{i}^{d}.{\bf S}_{i}^{c}.
\end{equation}
There are two major issues that are of interest from both
fundamental and experimental (heavy fermion materials) perspectives: (1.)
Description of the  competition between the Kondo effect and the RKKY interaction,
 which
generally  leads to a quantum phase transition from a paramagnetic to a
 long-range ordered phase, and (2.) The possible breakdown of the Kondo effect
(screening) near such
 a phase boundary, especially in antiferromagnetic metals \cite{Piers}. 
 We concentrate mostly on the first aspect, and consider the model
 Eq.(1) on a square lattice at finite hole doping $\delta$ ($\delta=0$ is
 half-filling), which makes the system conducting.  

We rewrite Eq.(1) in the local (one site) basis, where four bosonic
 modes (a singlet and a triplet) can be excited: 
$\sqrt{2}s^{\dagger}|v \rangle =\epsilon_{\sigma \rho}c_{\sigma}^{\dagger}
d_{\rho}^{\dagger}|0 \rangle$,
 $\sqrt{2}t_{\alpha}^{\dagger}|v \rangle = \sigma_{\rho \delta}^{\alpha} 
\epsilon_{\delta \sigma}c_{\sigma}^{\dagger}d_{\rho}^{\dagger}
|0 \rangle$, $(\alpha=x,y,z; \ \sigma,\rho,\delta=\uparrow,\downarrow)$,
 as well as two fermionic modes: 
$f_{1,\sigma}^{\dagger}|v \rangle = d_{\sigma}^{\dagger}|0 \rangle$,
$f_{2,\sigma}^{\dagger}|v \rangle = c_{\uparrow}^{\dagger}c_{\downarrow}^{\dagger}
d_{\sigma}^{\dagger}|0 \rangle$. Here $|v \rangle$ is a reference vacuum
 with no excitations present, and
$|0 \rangle$ corresponds to absence of physical electrons. The described set of
 states can be used to generalize  
  the bond-operator representation, often used in 
 localized spin systems (with only $s$ and ${\bf t}$ modes present) \cite{Subir}.
 Similar ideas have also been successfully applied to
  Kondo insulators ($\delta=0$) \cite{Bond}.
The representation is exact provided  the on-site  kinematic 
constraint is implemented:
$s^{\dagger}s + f_{\nu,\sigma}^{\dagger}f_{\nu,\sigma}+
t_{\alpha}^{\dagger}t_{\alpha} =1$, necessary to 
 restrict the Hilbert space of the new operators to the physical
sub-space. It is physically clear that
 for doping $\delta = 0$ and strong Kondo exchange $J/t\gg1$ the ground state
 is a singlet (thus paramagnet), meaning that the $s$ excitations condense: 
$\langle s \rangle \neq 0$. Let us note that  the term ``strong-coupling"
($J/t\gg1$) in this work has the opposite meaning
 to the one used in Ref.\cite{Piers}. 
 The Hamiltonian in the local representation  becomes, by taking the singlet
 state  $s^{\dagger}|v \rangle$ as the ground state, 
$H=H_{2}+H_{Int}$:   
\begin{equation}
\! \! H_{2} \!  = \! \! \!  \sum_{{\bf k},\sigma,\nu=1,2}E_{\nu}({\bf k})
f_{\nu,\sigma,{\bf k}}^{\dagger}
f_{\nu,\sigma,{\bf k}}
+ J \sum_{{\bf k}}{\bf t}_{{\bf k}}^{\dagger}.{\bf t}_{{\bf k}}+
\sum_{{\bf k},\sigma}B({\bf k})f_{1,\sigma, {\bf k}}^{\dagger}
f_{2,-\sigma,-{\bf k}}^{\dagger}
+ \mbox{h.c.},   
\end{equation}
\begin{equation}
 H_{Int} =  \  \sum \ [{\bf t}{\bf t}ff] \ \& \ [{\bf t}ff]  \ \mbox{scattering} \ \mbox{vertices},
\end{equation}
where $E_{\nu}({\bf k}), B({\bf k}) \propto t$ are easily calculated,
 and $H_{Int}$ represents the non-linear interactions
 (with coefficients also proportional to the hopping $t$) between the
bosonic and fermionic modes. The challenge  now becomes to treat the kinematic 
constraint (which amounts to an infinite local repulsion between the ${\bf t}$
 and $f_{\nu}$
modes),  as well as the interactions present in $H_{Int}$, as accurately
 as possible in order to have reliable results for the excitation spectrum.
 We have followed the philosophy outlined in Ref.\cite{Us} for spin insulators
 to  implement the hard-core repulsion, while the various terms in $H_{Int}$
 were treated to lowest non-trivial (one-loop) order  
 (see also references \cite{Oleg} and \cite{Park}, discussing the situation at finite doping
for the $t-J$ model). The extension of the approach from zero to finite doping is
 a non-trivial matter, and details will be provided elsewhere \cite{Us1}.
 It is crucial in this treatment that the S=1  magnetic excitations (the spectrum of
 the ${\bf t}$ operators) form a dilute Bose gas. Departure from the dilute,
 low-density limit would also mean a possible change in the ground state 
 and failure of  our approximation scheme since the 
latter is  based on classification of
 diagrams in powers of the excitation density. In addition, doping is also
 assumed to be a small parameter in the problem.
   
Let us summarize our  main results \cite{Us1}: 
\vspace{0.05cm}
\newline
$\bullet$ The magnetic (${\bf t}$) excitations are found to be gapped everywhere in a
certain region
 of parameter space, see Fig.1, meaning that the system is paramagnetic. The
 gap at the antiferromagnetic ordering wave vector ${\bf Q}_{AF}=(\pi,\pi)$
 vanishes on a critical line, signaling a transition to a phase with long-range order
 (finite staggered magnetization, proportional to the condensate
$\langle {\bf t}_{{\bf k}={\bf Q}_{AF}} \rangle $).
 The parameter F, displayed on the axis in Fig.1 (which is introduced for
 numerical reasons), is found to vary very weakly with doping.
 In the simplest of our approximation schemes it has the constant value
 F$\approx 0.72$ \cite{Us1}.
 This produces a critical point at zero doping $(J/t)_{c} \approx 1.39$ which 
is quite close
 to the numerical result \cite{Numerics} (however no numerical results are
 available for $\delta \neq 0$). Let us also mention that
 the magnons are  damped at any finite doping,
 which can readily be seen from the width of their spectral function.
\vspace{0.1cm}  
\newline
$\bullet$ We would like to emphasize that our  calculational scheme
is self-consistent in the sense that the dilute Bose gas description is maintained
throughout the phase diagram. This means that the singlet ground state,
 the starting point of the calculation, is stable (i.e. $\langle s \rangle \neq 0$), 
 leading to the usual Fermi liquid description of the critical
 properties \cite{HM}. Departure from the Fermi liquid picture would
 occur if $\langle s \rangle  \rightarrow 0$ at the quantum critical point (QCP); 
one can then
 expect various exotic effects such as different critical exponents, which
 has been  in fact suggested to occur in two dimensions
 \cite{Piers}. The fact that we do not
 find perturbative indications for breakdown of the Kondo effect 
at the QCP could be quite
 possibly due to our
 approximation scheme, which  can detect the existence of 
a quantum critical line (where [the magnetic gap]$\ll J$), but  is not capable
 of predicting universal properties in the quantum critical regime. 

We gratefully acknowledge stimulating conversations with Kevin Ingersent and
 Subir Sachdev, and the financial support of NSF Grant
DMR-9974396.

%
%
%
%

\begin{figure}
\centering
\includegraphics{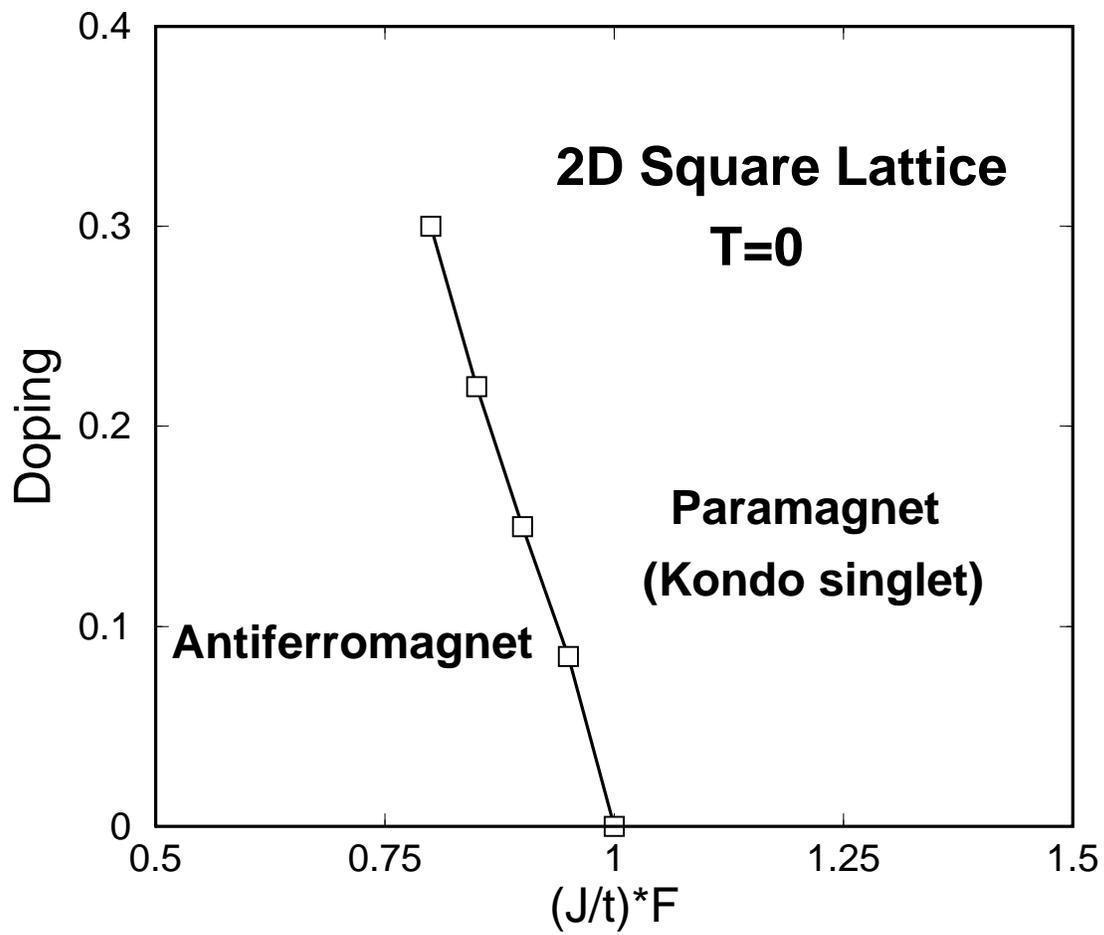}
\caption{Phase diagram of the Kondo lattice model.} 
\end{figure}

\end{document}